\documentclass{aa}

\usepackage{graphicx}
\usepackage[normalem]{ulem}
\usepackage{txfonts}
\usepackage{natbib}
\usepackage[colorlinks=true,citecolor=blue,linkcolor=blue]{hyperref}

\begin{document}

 \title{Opacity distribution functions for stellar spectra synthesis}

\author{M. Cernetic
    \inst{1}
    \and
    A. I. Shapiro\inst{1}
    \and
    V. Witzke\inst{1}
    \and
    N. A. Krivova\inst{1}
    \and
    S. K. Solanki\inst{1,}\inst{2}
    \and
    R. V. Tagirov\inst{1,}\inst{3}
    }

\institute{Max-Planck-Institut f\"ur Sonnensystemforschung, Justus-von-Liebig-Weg 3, 37077, G\"ottingen, Germany \\ \email{cernetic@mps.mpg.de}
     \and
     School of Space Research, Kyung Hee University, Yongin, Gyeonggi, 446-701, Republic of Korea
     \and
     Astrophysics Group, Imperial College London, London SW7 2AZ, UK
     }

\date{Received \textbf{???}; accepted \textbf{???}}

\abstract{Stellar spectra synthesis is essential for the characterization of potential planetary hosts. In addition, comprehensive stellar variability calculations with fast radiative transfer are needed  to disentangle planetary transits from stellar magnetically-driven variability. The planet-hunting space telescopes, such as CoRoT, Kepler, and TESS will bring vast quantities of data, rekindling the interest in fast calculations of the radiative transfer.} 
{We revisit the Opacity Distribution Functions (ODF) approach routinely applied to speedup stellar spectral synthesis. To achieve a considerable speedup relative to the current state-of-the-art, we further optimize the approach and search for the best ODF configuration. Furthermore, we generalize the ODF approach for fast calculations of flux in  various filters often used in stellar observations.}
{In a parameter-sweep-fashion, we generated ODF in the spectral range from UV to IR with different setups. The most accurate ODF configuration for each spectral interval was determined. For calculations of the radiative fluxes through filters we adapted the wavelength grid based on the transmission curve, whereafter the normal ODF procedure was performed.}
{Our optimum ODF configuration allows for a three fold speedup, compared to the previously used ODF configurations. The ODF generalization to calculate fluxes through filters results in a speedup of more than two orders of magnitude.}
{}
 


\maketitle
%
\section{Introduction}
\label{sec:intro}

Radiative transfer calculations over broad spectral intervals are important for modelling stellar and planetary atmospheres. One of the main challenges in such calculations is the treatment of a huge number of atomic and molecular lines typically found in a stellar or planetary spectrum. Atomic and molecular lines contribute immensely to the total opacity in the solar atmosphere, and even more in stars cooler than the Sun. The linelists that are currently employed for modelling and interpreting the solar spectrum contain more than hundred millions spectral lines \citep{kurucz2005}. To properly account for all these lines one has to perform calculations on a fine wavelength grid (typically with a resolution of just a few m\AA, which results in millions of grid points). This is computationally demanding.

Fortunately, many applications only need the knowledge of the amount of energy emitted in comparatively broad spectral intervals (e.g. a few \AA\ wide). Such applications include interpretation of solar observations with filter imagers \citep[see][and references therein]{solar_b_hinode_2008, sunrise_2017}, modelling solar and stellar brightness variations \citep{ermolli_2013, Solanki_Krivova_Solar_Irradiance_Variability_and_Climate_2013, yeo_rev_2016} and generally the analysis and interpretation of low-resolution spectra. The analysis low-resolution spectra will be even more important with the 3rd Gaia data release early 2021 providing hundreds of millions of low-resolution spectra \citep{gaia_2016}.

A straightforward approach for obtaining a low-resolution spectrum is to first calculate the spectrum at high-resolution directly accounting for all important lines, and then to convolve it with a given spectral response function \citep[see, e.g. Fig. 1 from][]{riethmueller_2014}. An obvious disadvantage of this approach is that calculations still have to be performed on a fine wavelength grid. An alternative to the straightforward approach is the employment of Opacity Distribution Functions \citep[ODF, see e.g.][pp.625-627]{hubeny_mihalas_2015}. ODF allow accounting for effects from an arbitrary number of spectral lines while calculations are done on a coarser wavelength grid. This is done by approximating the complex structure of opacity although with some loss of accuracy. In order to obtain the ODF, the entire spectral domain is split into bins. All wavelength information within a bin is lost and only the bin-averaged intensity values are calculated. For that, the line opacity in each of the bins is approximated by several, typically a dozen or less, representative values. The total opacity in the bin is then the sum of the line opacity described by ODF and the continuum opacity, for which the change within the bin is neglected.

In regions where local thermodynamic equilibrium (hereafter LTE) is valid, the opacity only depends on wavelength, the local conditions, namely the temperature and the pressure, and to some extent on the velocity distribution in the atmosphere. In 1D calculations, the velocity distribution is quantified by the microturbulence velocity. This allows the ODF to be pre-tabulated on a grid of temperatures and pressures. Interpolation is used to get the ODF values for a specific temperature and pressure. The interpolated ODF are then used for radiative transfer, skipping all opacity calculations. In the non-LTE case, opacity depends on the non-local radiation field making it unsuitable for such a pre-tabulation. Several non-LTE investigations \citep[see, e.g., Sec.~6 in ][]{werner_dreizler_1999_classical} made use of pre-tabulated photon cross-sections instead of opacities.

The ODF approach had been actively developed until the late 80s \citep{labs_odf_1951, kurucz_model_atmospheres_1979, Anderson_1989}, and a number of algorithms for the optimal approximation of the opacity distributions within the bins were proposed. \cite{kurucz_model_atmospheres_1979} approximated ODF by 10 values paying special attention to the accurate sampling of the high opacity tail of the distribution. Later, \cite{castelli2004} used a modification of this approach employing 12 values for approximating opacity within the bin to synthesize spectra over a broad range of stellar fundamental parameters. In recent years ODF have not been further developed because computers were getting so fast that single 1D model calculations do not require any further acceleration. 

Recently 3D magnetohydrodynamic (MHD) simulations of solar and stellar atmospheres reached a new degree of realism \citep{stein_lr_2012, nordlund_lr_2009, beeck_ii_2015, beeck_i_2015}, which led to the development of the 1.5D radiative transfer approach. In the 1.5D approach, the emerging spectra are calculated along many rays passing through a 3D MHD cube. The resulting intensity in a given direction is then calculated by averaging the intensities along multiple parallel rays, which sample the whole simulation cube \citep{holzreuter-solanki-2012, holzreuter-solanki-2013, rietmueller-solanki-2014, norris-beck-2017}. For example, for a $512 \times 512$ MHD cube at 8 viewing angles radiative transfer calculations on more than 2 million (512*512*8) 1D atmospheres have to be performed. It is therefore crucial for the 1D calculations to be as fast as possible, providing the motivation for a further improvement in ODF. We note that while the 1.5D approach does not allow accounting for the effects of ``cross-talk'' between different rays it has been shown to yield an accurate results with exception of cores of very strong lines \citep{holzreuter-solanki-2013,PereiraundUitenbroek}
 
Thus, in this paper we build on previous work regarding the ODF and search for an optimal setup for the desired trade-off between speed and accuracy in the case of the quiet sun model atmosphere. The qualitative conclusions are also valid for other stars. In addition, the ODF approach is extended to intensity calculations through broadband filters.

This paper is structured as follows. In Section~\ref{sec:problem_description_and_method} we give a detailed description of ODF and the setup used to perform this investigation. In Section~\ref{sec:odf_and_falc} we analyze the performance of ODF using an example atmosphere model for different sub-bin sizes and different sub-bin combinations. Moreover, we investigate how the optimal choice of sub-bins changes for different spectral intervals. Finally, a discussion and our conclusion are presented in Section~\ref{sec:conclusions}.

\section{Problem description and method}
\label{sec:problem_description_and_method}
\begin{figure}[h]
	\centering
	\includegraphics[width=\columnwidth]{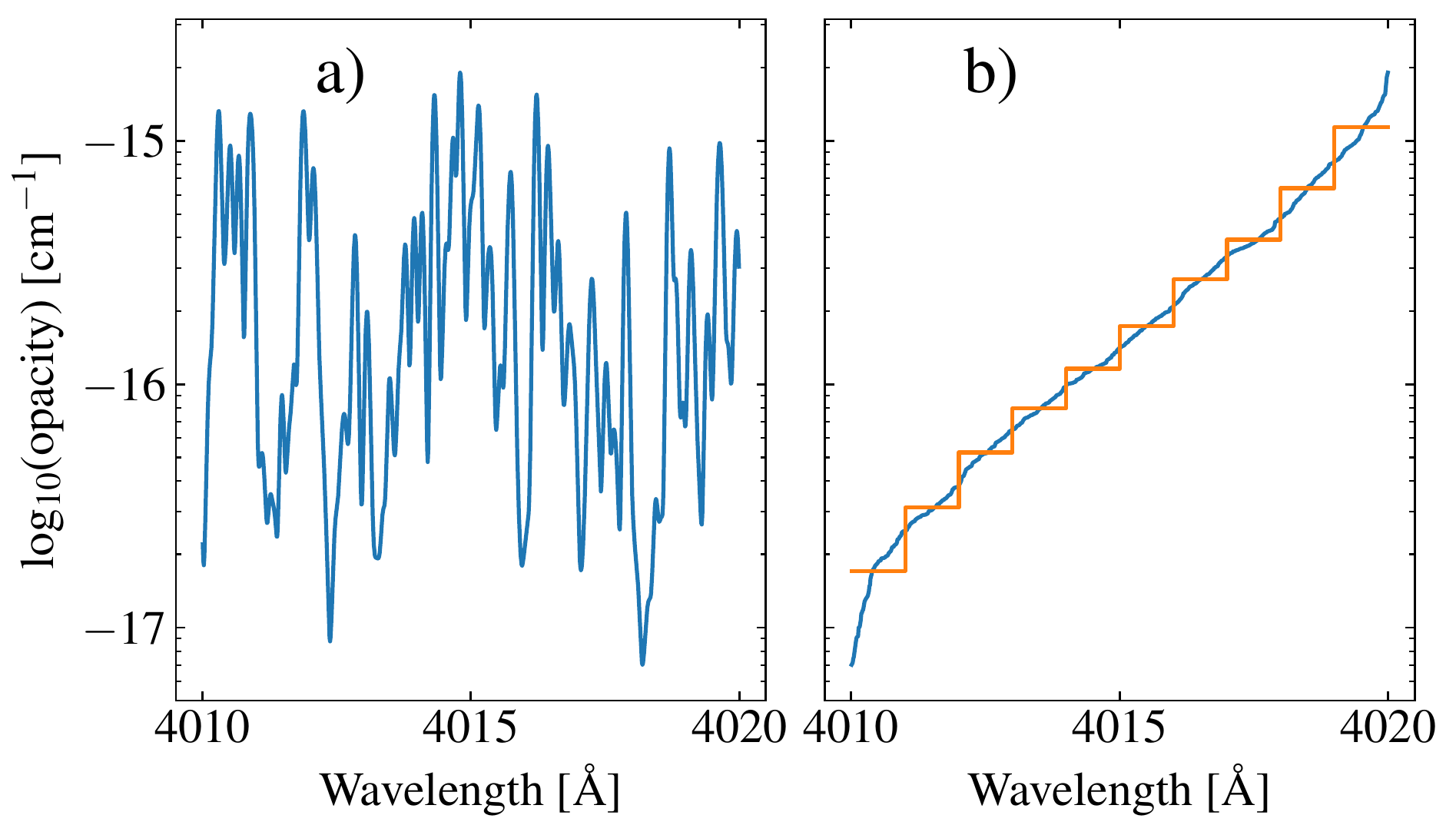}
    \caption{Opacity variation within 10~\AA\ (panel a) as well as sorted opacity and the corresponding mean values over individual sub-bins (blue and orange in panel b, respectively).}
    \label{fig:sub_bin_values}
\end{figure}

The opacity in stellar atmospheres has a complicated wavelength dependence in particular, in spectral lines, changes reach up to several orders of magnitude within a narrow wavelength interval, as illustrated in Fig.~\ref{fig:sub_bin_values}a. As a result, opacity calculations require a fine wavelength grid. While some applications need calculations of high-resolution spectra, for many applications low-resolution spectra are sufficient. Such low-resolution spectra can be obtained using accelerated methods.

There exist two conceptionally different methods to approximate low-resolution opacities, opacity sampling (OS) and opacity distribution functions (ODF) \citep[pp.~625-627]{hubeny_mihalas_2015}. With OS the high resolution opacity is simply re-sampled at a lower resolution. It is crucial that the sample preserves the original statistics of the high resolution opacity. ODF, introduced in Sect.~\ref{sec:intro}, allow achieving a similar level of approximation but with fewer representative opacity values. This work solely focuses on the ODF.

To calculate ODF, the high-resolution wavelength grid is divided into bins. In each bin the opacities are sorted in an ascending order as illustrated in Fig.~\ref{fig:sub_bin_values}b. Each bin is further divided into sub-bins and the mean value of the opacities is obtained for each of the sub-bins. Figure~\ref{fig:sub_bin_values}b shows, as an example, one 10\,\AA\ wide bin, subdivided into 10 sub-bins. Panel Fig.~\ref{fig:sub_bin_values}a shows the original distribution of opacities on a high-resolution grid, while panel Fig.~\ref{fig:sub_bin_values}b illustrates the ODF.

All calculations and technical implementation presented in this paper are performed with the Non local thermodynamic Equilibrium Spectral SYnthesis Code \citep[NESSY,][]{tagirov_2017_nessy}, which is a further development of the COde for Solar Irradiance \citep[COSI,][]{shapiroetal2010}. NESSY takes an atmosphere model, abundances, and linelists for elements from hydrogen to zinc as input and calculates emergent spectra in the range from UV to radio. All spectra in this work are disk-integrated, i.e. corresponding to radiative flux.

We added support for ODF in the spectral synthesis part of NESSY.
In the original version, first the Saha-Bolzmann equation is used to compute the electron level populations and then opacities from them. Then the radiative transfer (RT) is solved on a fine wavelength grid. In the new version, we first synthesize ODF from the high-resolution opacity and then solve the radiative transfer equation for each of the ODF sub-bins. Fluxes from individual sub-bins are summed together to get the flux of the whole bin
\begin{equation}
    \mathcal { F } = \sum_{i=1}^{n} \mathcal { F }_i \frac{\Delta \lambda_i}{\Delta \lambda},
    \label{eq:sub_bin_weight}
\end{equation}
where $\mathcal { F }_i$ and $\Delta \lambda_i$ are the flux from and the width of the sub-bin respectively, while $\Delta \lambda$ is the width of the bin.
Whereas the fine wavelength grid shown in Fig.~\ref{fig:sub_bin_values}a) has 2000 points per nm, the  step-wise grid in Fig.~\ref{fig:sub_bin_values}b) has only 10 points per nm. This results in a 200-fold faster RT calculation. We note that this is only the speedup for the spectral synthesis calculations. The RT calculations take most of the computational time, since the generation of ODF has to be performed only once for the given ODF configuration, linelist, and abundances. Thus we are not considering any speedup in the ODF generation itself. In the next section we investigate how accurate and fast the ODF approximation depends on the size of the sub-bins and their distribution within the bins.

\section{Optimal ODF for spectral synthesis}
\label{sec:odf_and_falc}
An optimal ODF setup for spectral synthesis is a compromise between the speed of the computations and the accuracy of the result, which also depends on a specific task. In this section, we analyze how the resulting accuracy depends on the size and the distribution of the sub-bins. For this, we use, as an example, the FALC atmospheric structure of the quiet Sun from \cite{fontenlaetal2006}. The molecular and atomic linelists used in NESSY have been taken from \cite{shapiroetal2010}. To analyze the performance of ODF, we take the flux computed with various ODF setups (i.e. on a coarse wavelength grid, hereafter ODF spectra). Then we take the flux computed on a fine wavelength grid and average it within each ODF bin (hereafter detailed spectra). Such a detailed spectrum represents the case when the line opacity is taken into account without any approximation. We then compare ODF and detailed spectra to study the performance of various ODF configurations.

\subsection{Mean opacity}
\label{sub_sec:mean_opacity}
To illustrate the power of the ODF approach we start by comparing its performance to a straight-forward method of employing mean opacities (i.e. ODF using only one sub-bin).

\begin{figure}
    \centering
    \includegraphics[width=\columnwidth]{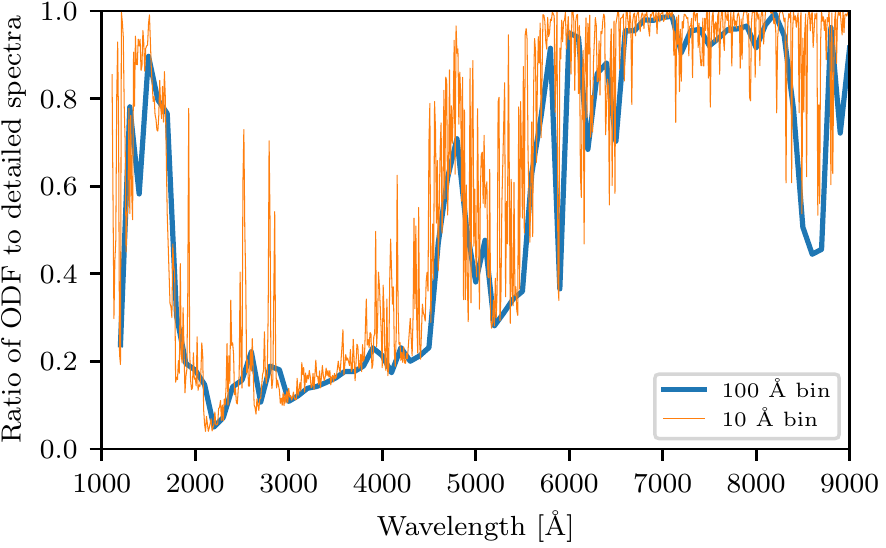}
    \caption{Ratio of spectra calculated using mean opacity to detailed spectra. Bin size of 10\,\AA\ (orange) and bin size of 100\,\AA\ (blue).}
    \label{fig:ODF_vs_nonODF}
\end{figure}

We perform these calculations for two bin sizes, 10\,\AA\ and 100\,\AA. Figure~\ref{fig:ODF_vs_nonODF} illustrates that calculations using mean opacities cannot adequately reproduce the spectrum, for bin sizes somewhat greater than the high-resolution grid with a resolution of 200 points per \AA. In particular, they lead to the underestimation of the intensity by up to an order of magnitude in the UV part of the spectrum. This can be explained by the wide range of opacity values within the bins (see Fig.~\ref{fig:sub_bin_values}a) and the fact that a mean value is skewed by the extreme values in the sample. A large amount of photons escape the atmosphere at wavelengths where the opacity is low. When replacing opacity there by the higher mean opacity it becomes more difficult for photons to escape, and thus the emergent flux decreases (in other words averaging closes ``opacity windows''). 

Figure~\ref{fig:ODF_vs_nonODF} clearly indicates that a better solution than the mean opacity is needed. This can be achieved by sorting the opacity within each bin and then splitting the bin into several sub-bins, individually averaging the values within each of the sub-bins. Such a procedure prevents very high opacity values from skewing the average of other sub-bins, as it groups opacity with similar values together. 

\subsection{Uniform sub-bins}
\label{sec:uniform_sub_bins}
In this section we investigate the effect of number of sub-bins on the accuracy of the spectrum calculations. We first consider the case of sub-bins of uniform width.

\begin{figure}
    \centering
    \includegraphics[width=\columnwidth]{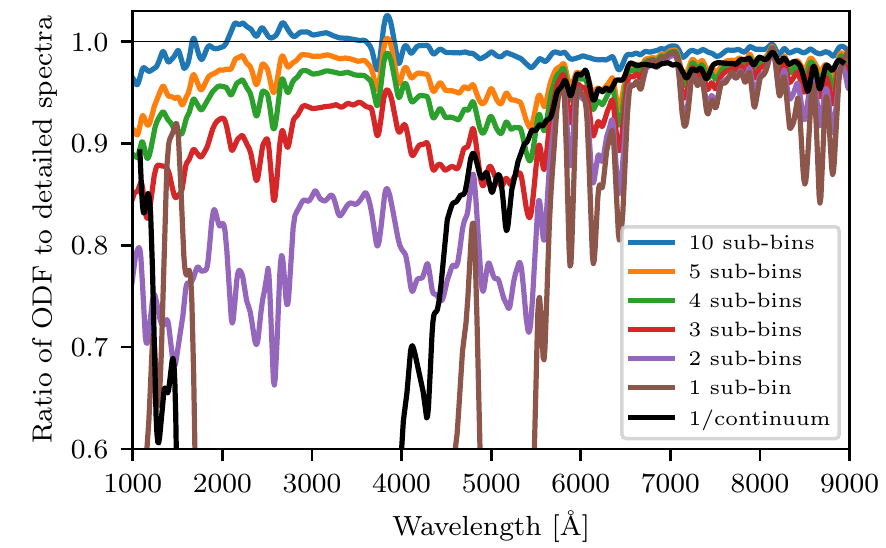}
    \caption{Ratio of ODF based spectra to the detailed one for different uniform sub-bin sizes, as given by the legend in the plot. The black curve is calculated with continuum opacity only. It is plotted on an inverted scale to allow for a better comparison. Bin size is 10\,\AA. Convolved with a Gaussian, $\sigma = 30\,\AA$.}
    \label{fig:1to5}
\end{figure}

Calculations with different numbers of sub-bins are presented in Fig.~\ref{fig:1to5}. As expected, an increase in the sub-bin number improves the accuracy of spectral synthesis, but this comes at the cost of the computational time. Comparing the brown and violet lines in Fig.~ \ref{fig:1to5}, one can see that the introduction of only one additional sub-bin (i.e. splitting the bin into 2 sub-bins) significantly improves the accuracy of the calculations. 
We additionally show the continuum only spectra (black line) to illustrate the total effect of spectral lines on the emergent spectrum. The simultaneous comparison between the continuum-,  ODF-, and detailed spectra- calculations shows how well the ODF approach accounts for the effect of spectral lines. It is worth noting the continuum spectrum provides a better approximation of the emergent spectrum than calculations with 1--5 uniform sub-bins longward about 6000\,\AA\ so that  including lines with a suboptimal ODF configuration might be worse than not treating lines at all.
Note that the black curve has been inverted around unity (continuum-only computations overestimate the emergent flux, while uniform ODF typically underestimate it). The uniform ODF in general allow adequate treatment of the huge amount of spectral lines in the UV, where lines completely dominate the continuum, forming the UV line haze. At the same time, the accuracy of the calculations remains low longward of about 5000\,\AA\ compared to the continuum only. Interestingly, more than five uniform sub-bins are needed to reach better accuracy compared to neglecting lines completely (i.e. taking only continuum opacity into account). The reason for such behavior is the small number of spectral lines longward of 5000\,\AA. In most bins only a small fraction of wavelengths have opacity with values much larger than that of the continuum. Only those opacities can noticeably affect the intensity. If the width of the last sub-bin (i.e. the sub-bin containing the greatest opacity values) is larger than this fraction of high opacities, the calculation becomes erroneous. The very high values skew the mean opacity of that sub-bin towards larger values, which results in inaccuracy in the intensity calculations. Another way to understand this is to consider an imaginary case where one line in the bin is significantly stronger than all other lines. If the width of the last sub-bin is larger than the width of this line, then the mean opacity in this sub-bin will be skewed towards larger values, introducing an error, which increases with the size of the last sub-bin. We note that the same effect, albeit amplified by the fact that it appears over the whole bin and not only in the last sub-bin, causes the inaccuracy in the calculations using only mean opacity values (see Sect.~\ref{sub_sec:mean_opacity}). The solution is to consider non-uniform sub-bins and pay special attention to the relative size of the last sub-bin.

\subsection{Non-uniform sub-bins}
\label{sec:nonuniform_sub_bins}
In this section we extend the analysis done in Sect.~\ref{sec:uniform_sub_bins} by considering non-uniform sub-bins.
We note that non-uniform sub-bins have already been employed by \cite{kurucz_first_odf_1974} and more recently by \cite{castelli2004}, who used an improved version of ODF with 12 sub-bins. They used the following relative widths of the sub-bins in Eq.~\eqref{eq:sub_bin_weight}: 1/10 for first nine sub-bins (i.e. sub-bins with lowest opacity values) and 1/20, 1/30, and 1/60 for the 10th, 11th, and 12th sub-bin respectively , with the highest opacity values. For our sub-bin combinations we also use an ascending order of opacity.

\begin{figure}
    \centering
    \includegraphics[width=\columnwidth]{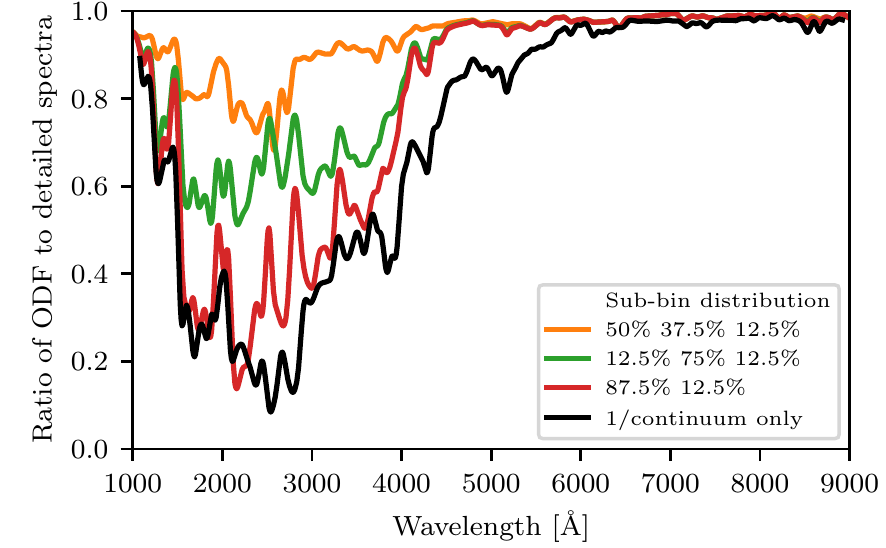}
    \caption{Ratio of ODF-based spectra using non-uniform bins to the detailed one. Legend shows sub-bin widths in parts of the bin size for each colored curve. The averaged value of opacity in the sub-bins increases from left to right. The black curve is calculated with continuum opacity only, inverted to allow for a better comparison. Bin size is 10\,\AA. Convolved with a Gaussian, $\sigma = 30\,\AA$.}
    \label{fig:non_uniform}
\end{figure}

Figure~\ref{fig:non_uniform} shows the ratio of the spectra computed for three different sub-bin distributions to the detailed spectrum. The last sub-bin is kept at 12.5\% of the total bin size. One of the sub-bin distributions (red line in Fig.~\ref{fig:non_uniform}) attributes all the remaining 87.5\% of opacities to a single sub-bin, while the other two (orange and green) split them between two additional sub-bins in two different ways. One can see that all three sub-bin distributions return roughly the same accuracy longward of 4500\,\AA, even the one with just two sub-bins. This is in line with the discussion in Sect.~\ref{sec:uniform_sub_bins}. The last sub-bin is able to account for all strong lines whose opacity should not be mixed with the opacities of the other lines. At the same time the opacity of the remaining weak lines affects the intensity linearly to first order so that it can be averaged without losing accuracy. All in all, just two sub-bins (albeit with different sizes) are sufficient for achieving an accuracy of 95\% longward 4500\,\AA. This finding is very important for modelling stellar brightness variations since they are often observed in the visible and infrared parts of the spectrum, e.g. the Kepler telescope measures brightness longward of 4000\,\AA.

For all cases, the accuracy significantly drops in the UV part of the spectrum, where the last sub-bin alone is insufficient to account for the larger range of opacities affecting intensity in a non-linear way. The worst performance of the considered sub-bin distributions is between 2000 and 3000\,\AA.

\begin{figure}
    \centering
    \includegraphics[width=\columnwidth]{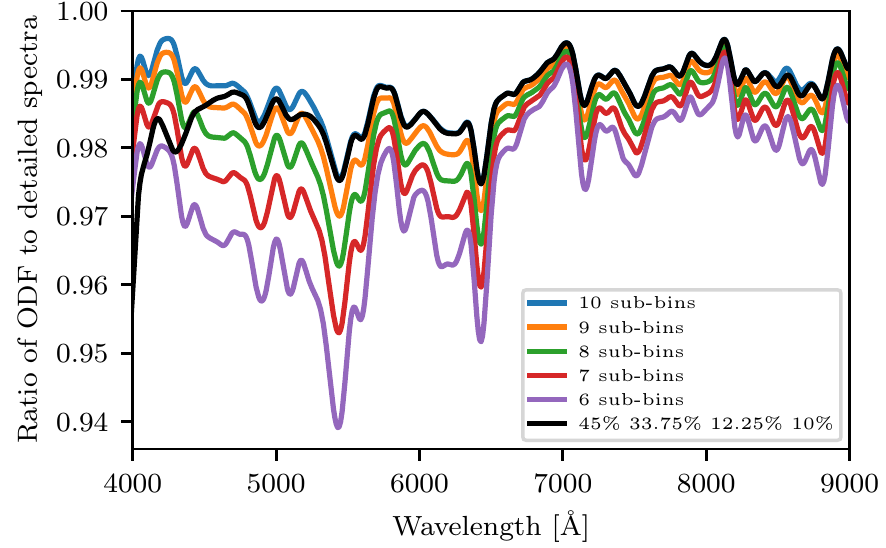}
    \caption{Ratio of ODF based spectra to the detailed one. The colored curves are for uniform sub-bins (see legend for the description), while the black curve shows the result of using just four non-uniform sub-bins. Bin size is 10\,\AA. Convolved with a Gaussian, $\sigma = 30\,\AA$.}
    \label{fig:uniform_vs_best_4_non}
\end{figure}

In Figure~\ref{fig:uniform_vs_best_4_non} we compare a non-uniform distribution with four sub-bins to uniform distributions with six and more sub-bins. The last sub-bin of the non-uniform distribution is the same size as the sub-bins of the uniform distribution with 10 sub-bins. Because of this the non-uniform distribution results in a high accuracy with fewer sub-bins longward of 4200\,\AA. Shortward of 4200\,\AA\ the non-uniform sub-bin distribution is not as accurate. The superior performance of the non-uniform sub-binning in the visible and infrared is promising, and below we study it more closely.

\subsection{Best combination of 4 and 10 sub-bins}
\label{sec:best_combination_finder}

\begin{figure*}
    \centering
    \includegraphics[width=180mm]{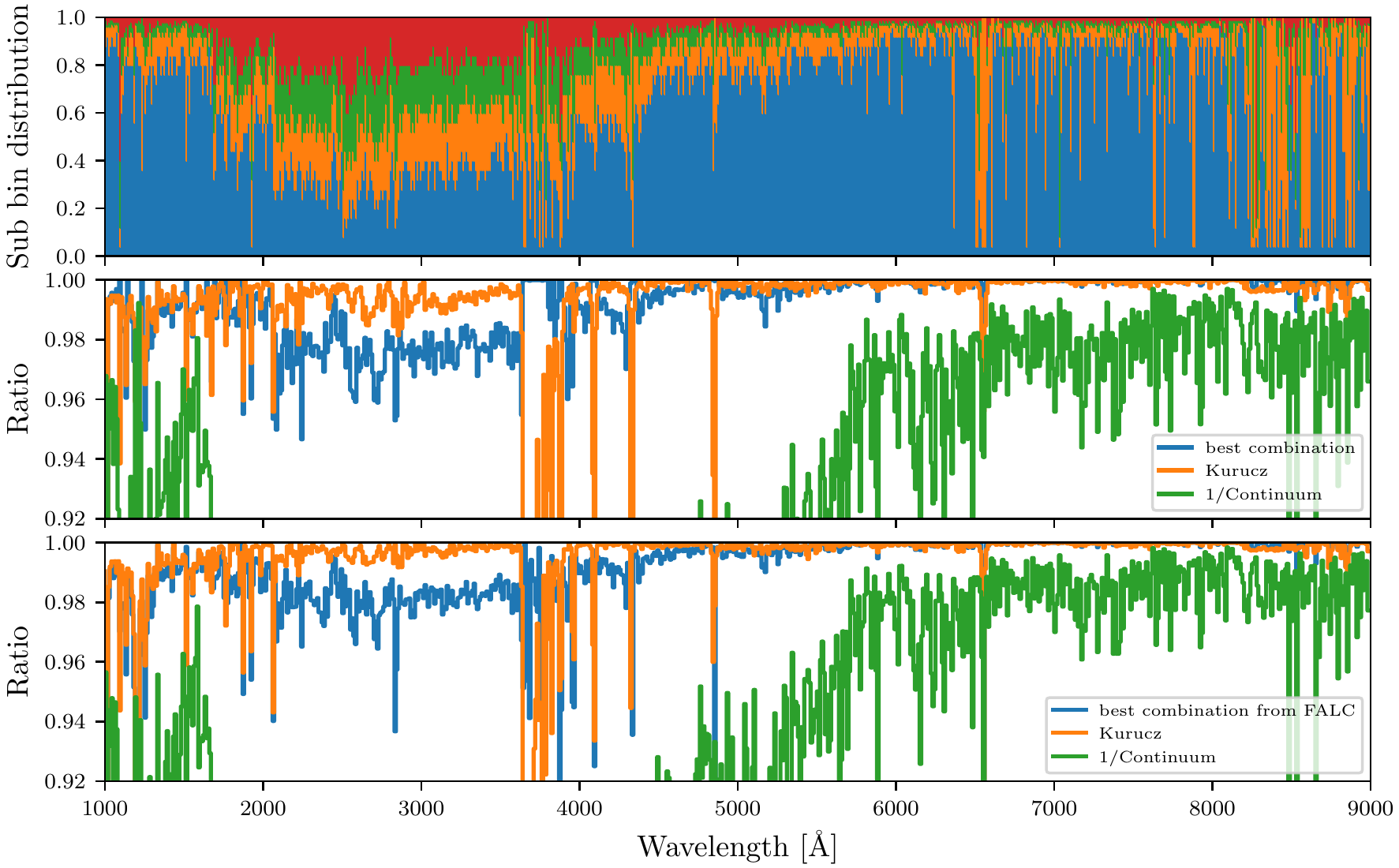}
    \caption{Results of the optimal sub-bin combination search with 4 sub-bins. Top panel: The optimal 4 sub-bin configuration of the 10\,\AA\ wide bins as a function of wavelength for the FALC atmosphere model. The color shows how sub-bins are distributed relative to the bin length. E.g. a column around 3500\,\AA\ can be interpreted as follows: the first sub-bin starts at the beginning of the bin and stretches up to 35\%, second ends at 60\%, third at 80\% and fourth at the end of the bin. Since opacity in the bin is sorted by the value, the colors also tell which opacity values fall in a sub-bin. The 35th percentile of opacity falls in the first sub-bin, opacity values from the 35th to the 60th percentile fall in the second sub-bin and so forth. Middle panel: Ratio of ODF to detailed spectra using the best combination of sub-bins for each bin from the upper panel (blue). For comparison we also show accuracy using Kurucz sub-bins (orange). Bottom panel: Ratio of ODF to detailed spectra using the best sub-bins for each bin from the upper panel applied to the FALP atmosphere model (blue) and using Kurucz sub-bins on the FALP atmosphere model (orange). The values of the ratio on the middle and bottom panels greater than one were mirrored over the abscissa y=1.}
    \label{fig:best_combination_finder}
\end{figure*}

In the previous two sections we showed that different wavelength domains require vastly different sub-bining to achieve high accuracy. We also demonstrated that uniform sub-bins worked relatively well in the UV, while non-uniform sub-bins were better for the IR. To investigate this further, we calculated the ODF spectra for all possible combinations of four sub-bins in the range from 1000\,\AA\ to 9000\,\AA\ with 10\,\AA\ bins. For this we have considered 28 possible locations of sub-bin borders within each bin. The number 28 was chosen as a good trade-off between the number of sub-bins that have to be considered: 23 possible locations of sub-bin borders from 4\% to 92\% of the bin size (with a uniform step of 4\% of the bin size) and five additional borders at 94\%, 96\%, 97\%, 98\%, 99\% of the bin size.
Each selection of 3 borders from the 28 possible values corresponds to one sub-binning so that there are $\binom{28}{3}$~=~3276 possible sub-bin combinations. We calculated fluxes for each of them and searched for the best sub-bin combination depending on the wavelength. Results  are presented in Fig.~\ref{fig:best_combination_finder}. The upper panel shows the best sub-bin distribution for each of the 10\,\AA\ bins. It confirms our previous findings that the UV part of the spectra requires approximately uniform sub-bins, while a small last sub-bin works best longward of~$\sim$4500\,\AA. 

Various model atmospheres differ in structure and it is not obvious that an optimal sub-bin combination found for one structure would lead to adequate results for another. To test this, we considered two atmosphere models by \cite{fontenlaetal2006}, FALC (model of the quiet sun) and FALP (model of the faculae). An optimal sub-bin combination that was found for FALC was then used to calculate the ODF spectra for FALP.
The lower panel of Fig.~\ref{fig:best_combination_finder} thus shows how the FALC optimal sub-bin combination performs on the FALP atmosphere. With a few exceptions around strong lines, the ODF reproduce the detailed spectral computations even better, mainly due to the weaker spectral lines in the FALP model.

The middle and lower panels in Fig.~\ref{fig:best_combination_finder} together show that using 3 times fewer sub-bins than in the recent setup by Kurucz we can achieve the same accuracy longward of 4500\,\AA. The accuracy of the 4-bin setup below 4500\,\AA\ is somewhat lower than but still comparable to that of Kurucz. Such optimized ODF are very valuable for applications where time is the main constraint. Note that our simple test suggests that the choice of the sub-bins can be frozen for typical solar atmospheres, so that the optimization needs to be done only once, tested for a few extreme atmospheres that may appear in a solar MHD simulation. After that they can be applied unchanged for solar applications (with the possible exception of sunspots, which we have not tested so far).

\begin{figure}
    \centering
    \includegraphics[width=\columnwidth]{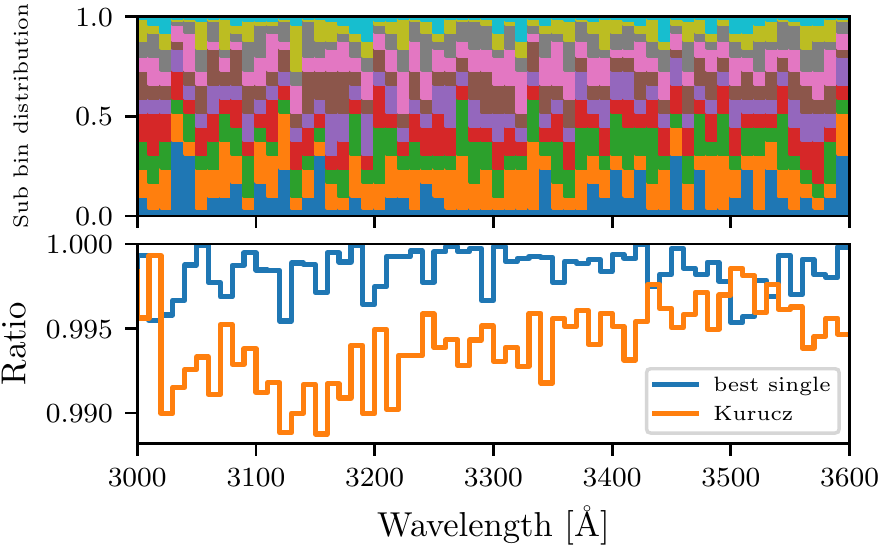}
    \caption{Results of the optimal sub-bin combination search with 10 sub-bins. Top panel: shows the best sub-bin distribution for each bin as a function of wavelength for the FALC atmosphere model. For the description of colors confer Fig.~\ref{fig:best_combination_finder}. Bottom panel: ratio of ODF to detailed spectra showing the single unique sub-bin combination with the lowest cumulative least squares error in all bins (green), using Kurucz sub-bins (orange). The curve showing best sub-bins for each bin from the upper panel has such a low error it would be indistinguishable from the y-axis.}
    \label{fig:best_combination_finder_10}
\end{figure}

Next we address applications where very high accuracy and some acceleration are important. This is achieved by increasing the number of sub-bins to ten. And more importantly, having 2.5 times more sub-bins that we had in the first case should bring considerably better accuracy. To make the number of possible sub-bin combinations suitable for the computing power at hand, we split the bin more finely than before. Possible sub-bin borders are now set at 4\%, 10\%, 17\%, 24\%, 31\%, 38\%, 45\%, 52\%, 59\%, 66\%, 73\%, 80\%, 84\%, 88\%, 92\%, 96\%, 98\%, 99\%. This results in $\binom{18}{9}$~=~48620 possible sub-bin combinations. To  reduce the computation time we only considered a narrow wavelength range, namely from 3000\,\AA\ to 3600\,\AA\ we choose this range because it is one of the trickiest spectral domains for ODF (see, e.g. Fig.~\ref{fig:best_combination_finder_10}).

Result using ten sub-bins are plotted in Fig.~\ref{fig:best_combination_finder_10}. The upper panel shows the optimal sub-bin combination for each bin. Compared to the upper panel of Fig.~\ref{fig:best_combination_finder} it is seemingly random. The reason is the large number of sub-bin distributions considered for each bin, and typically there is more than one possible sub-bin combination that allows high accuracy. This is further confirmed in the lower panel. The blue line shows the performance of the single unique sub-bin combination, that is one sub-bin combination for all bins, with the lowest cumulative least squares error in all bins. The boundaries of the sub-bins have been kept fixed at 4\%, 10\%, 17\%, 38\%, 52\%, 66\%, 80\%, 84\%, 88\%. The single unique sub-bin combination with the lowest cumulative least squares error in all bins performs better than Kurucz standard sub-bins plotted in green. This shows that our method of optimizing ODF can be tuned to provide very high accuracy but still provide a 20\% speedup.

\subsection{Different bin sizes}
\begin{figure}
	\centering
    \includegraphics[width=\columnwidth]{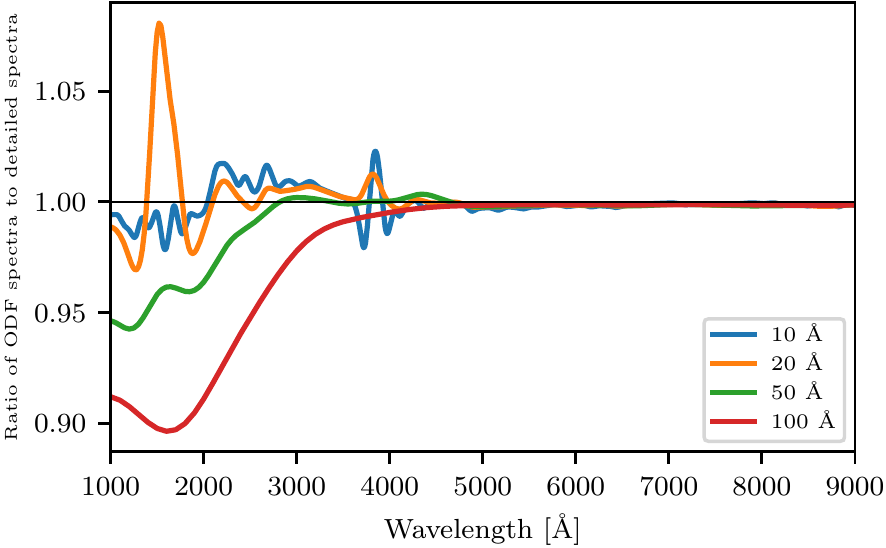}
    \caption{Ratio of ODF to detailed spectra with different bin sizes, as listed in the legend. Kurucz sub-bins. Smoothed using a Gaussian.}
    \label{fig:bin_size_comparison}
\end{figure}

In previous sections we investigated how to speedup calculations by decreasing the number of sub-bins per bin to either 4 or 10 sub-bins. Another possible approach is to keep a fixed number of sub-bins per bin but increase the bin size, thereby reducing the number of bins for a given  wavelength range, although at the cost of lower spectral resolution. ODF performance using Kurucz sub-bins with different bin sizes is presented in Fig.~\ref{fig:bin_size_comparison}.
    
Above about 5200\,\AA, the accuracy of the calculations barely depends on the bin size. This is important for applications that deal with instruments observing in that regime, like Kepler or TESS, as it allows us to use broad bins. Below 5000\,\AA, however, an increase in the size of the bins leads to poorer accuracy. This is because the large number of UV lines cannot be sufficiently treated by such ODF. Knowing the behaviour of ODF in larger bins is useful for calculations of spectra in, for example, the Kepler band. Treating such broad bands will be considered in the next section.

\subsection{Fluxes through filters}
\label{sec:filter_calculations}
Broad-band filter systems (e.g. UBV photometric system) have been a staple of stellar astronomy for a long time. The interest in accurate calculations of fluxes through filters is being rekindled by the advent of planetary hunting missions like Kepler, TESS, PLATO and others. In this section we introduce a more efficient method for computing fluxes through filters employing ODF.

The filter is characterized by the transmission curve, which describes the fraction of the transmitted spectral flux as a function of wavelength. The direct way of calculating flux in a given filter is to use the detailed opacity, calculate the detailed flux $\mathcal{F}_\lambda$, then multiply it with the transmission curve $\mathcal{C}_\lambda$, and finally integrate over wavelengths: 

\begin{equation}
    \mathcal { F } = \int_{\lambda_{L}}^{\lambda_{R}} \mathcal { F } _ { \lambda } \mathcal{C}_\lambda\  d\lambda,
    \label{eq:filter_flux}
\end{equation}
where the integral runs over the filter passband spanning from $\lambda_L$ to $\lambda_R$. 

The high-resolution opacity must be computed to find $\mathcal { F } _ { \lambda }$ so that such a direct approach has exactly the same disadvantages as calculating the detailed spectra. At the same time the application of the ODF is not straightforward since we cannot account for the filter wavelength dependence within the bin. In principle, this can be circumvented by making bins sufficiently small, which is not optimal because of the increase in computational time.

A way around this is to incorporate the effect of the transmission function into the wavelength grid before the ODF procedure is applied. Namely, by transforming the wavelength grid using the following substitution:
\vspace{2em}
\begin{equation}
    \Tilde{\lambda} (\lambda^*) = \int_{\lambda_L}^{\lambda^*} \mathcal{C}_\lambda\ d\lambda .
    \label{eq:lambda_substitution}
\end{equation}
Flux ${\mathcal{F}}$ is then  obtained by the integration over the new wavelength grid as 
\begin{equation}
    {\mathcal{F}} = \int_{\Tilde{\lambda_L}}^{\Tilde{\lambda_R}} F_{\Tilde{\lambda}}\ d\Tilde{\lambda},
    \label{eq:filter_flux_new}
\end{equation}
where $\Tilde{\lambda_L} = \lambda_L$ and $\Tilde{\lambda_R} = \Tilde{\lambda}\left( \lambda_R \right) = \int_{\lambda_L}^{\lambda_R} C_\lambda\ d\lambda$.

\begin{figure}
	\centering
    \includegraphics[width=\columnwidth]{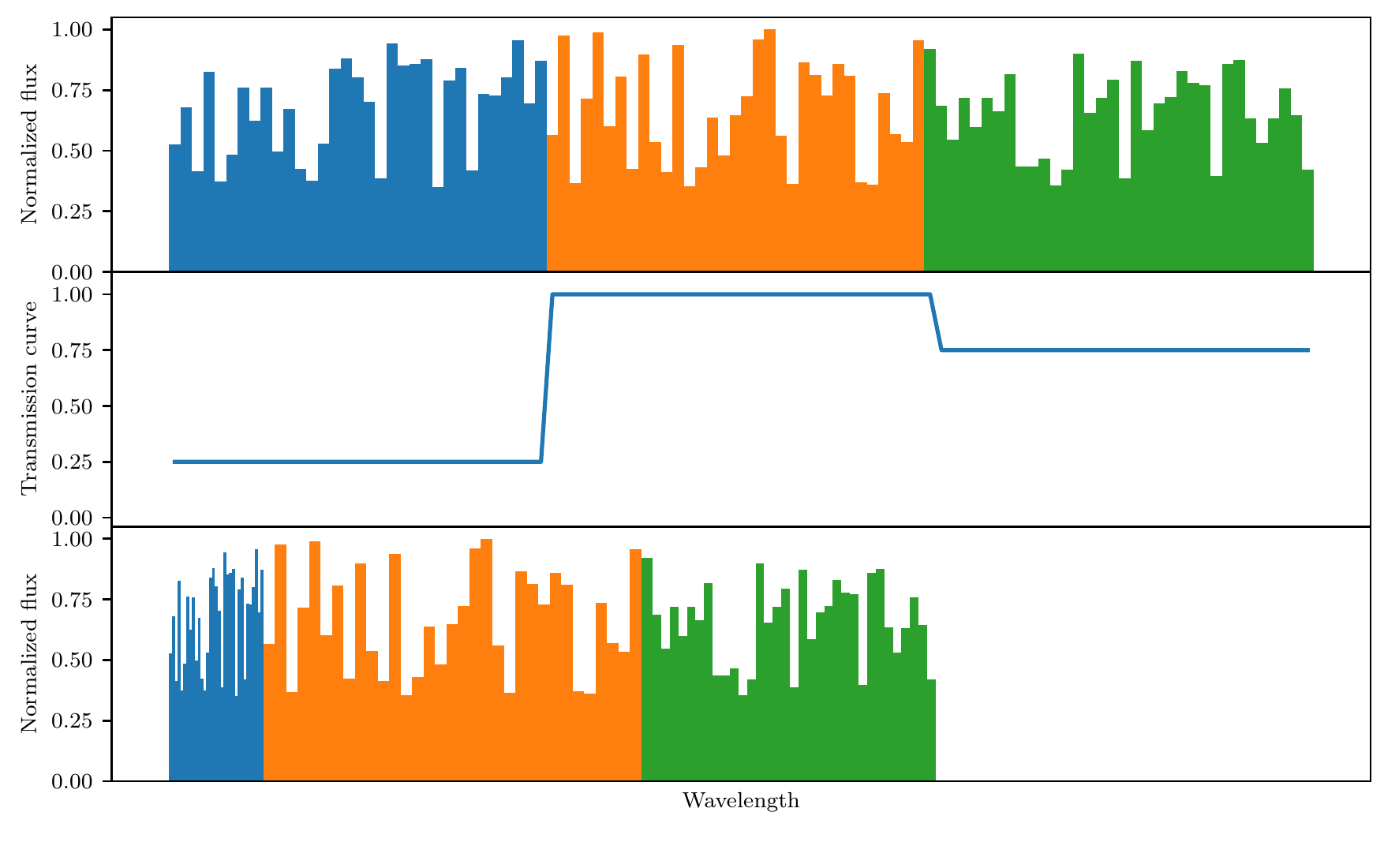}
    \caption{Illustration of the wavelength transformation according to Eq.~\eqref{eq:lambda_substitution}. Upper panel: schematic representation of flux dependence on wavelengths. Middle panel: exemplary step-wise transmission curve. Lower panel: resulting detailed flux on a transformed wavelength grid. Colors indicate to which of the three values of the transmission curve the fluxes are associated to.}
    \label{fig:odf_filters}
\end{figure}

The described wavelength transformation is schematically illustrated in  Fig.~\ref{fig:odf_filters}. The height and width of each column represent the flux value and the corresponding wavelength range, respectively. Our method boils down to multiplying the column width with the corresponding value of the transmission curve (as in Eg.~\eqref{eq:filter_flux_new}), instead of multiplying the height (as in Eg.~\eqref{eq:filter_flux}).

The main advantage of Eq.~\eqref{eq:filter_flux_new} is that, in contrast to Eq.~\eqref{eq:filter_flux}, it does not explicitly contain the transmission curve, and consequently the ODF approach can be applied.  

\begin{figure}
	\centering
    \includegraphics[width=\columnwidth]{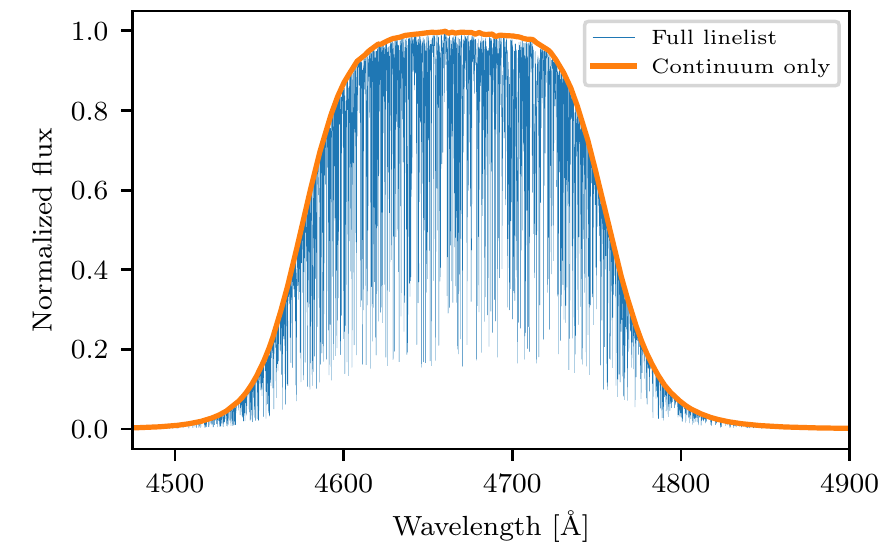}
    \caption{Normalized flux in Str\"omgren \textit{b} filter. Spectra synthesized using the full linelist (blue, plotted with decreased line width) and continuum only (orange).}
    \label{fig:b_filter}
\end{figure}

\begin{figure}
	\centering
    \includegraphics[width=\columnwidth]{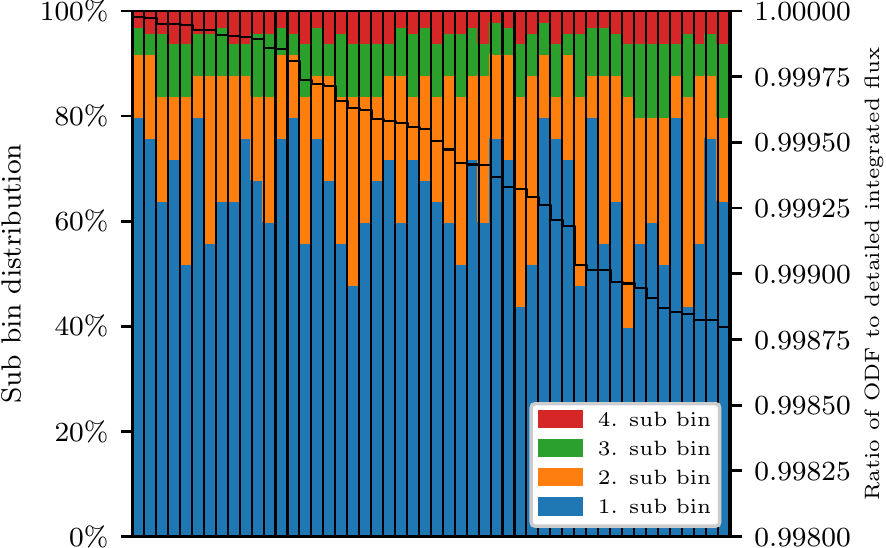}
    \caption{Performance of the ODF procedure for the Str\"omgren \textit{b} filter. Each column represents one sub-bin distribution. The color correspond to the left y-axis. For the description of colors confer Fig.~\ref{fig:best_combination_finder}. The distributions are sorted in a descending order according to the accuracy for the Str\"omgren \textit{b} filter shown by the black line and associated with the right y-axis.}
    \label{fig:optimal_stroemgren_1}
\end{figure}

In our investigation we applied such a filter ODF approach to the exemplary case of the  Str\"omgren \textit{b} filter. The filter transmission curve is plotted in Fig. 10 and has its effective central wavelength at 4666\,\AA \ with FWHM 156\,\AA. For numerical reasons we only consider wavelengths where the filter transmission curve is above 1\%, which is the interval from 4506\,\AA\ to 4836\,\AA, 330\,\AA\ in width.  We treat the entire 4506--4836 \AA \ range as one bin and search for the optimal configuration of sub-bins. The results are presented in Fig.~\ref{fig:optimal_stroemgren_1}.  One can see that  a very high level of accuracy is achievable with only 4 sub-bins. The best performing sub-bin distribution splits  the bin at 80\%, 92\%, and 97\% of bin size. We note that the accuracy of the filter ODF approach should be compared to the total contribution of lines (see Fig.~\ref{fig:b_filter}), which, according to our calculations, stands at approximately 15\%.

The speedup  we achieve by using 4 sub-bins compared to calculations using detailed opacity with a resolution of 80 points per \AA\ is $330 \times 80 / 4 = 6600$. To calculate the Str\"omgren \textit{b} flux with standard Kurucz ODF setup one needs 33 bins with 12 sub-bins, which results in 396 frequencies. Consequently, our approach gives us a speedup of roughly 100 times.

We note that in the ODF approach the Planck function change within a bin is neglected.  In the case of relatively narrow  Str\"omgren \textit{b} filter such a change can be neglected and thus the entire passband can be treated as one bin. This is not the case for broader filters, e.g. used in planetary-hunting missions. The application of the ODF approach to such filters will be considered in a forthcoming publication.

\section{Conclusions}
\label{sec:conclusions}
In this paper we have studied the optimal ODF setup for fast and reliable spectral synthesis. We find that the main factor determining the accuracy longward of 4500\,\AA\ is the width of the last sub-bin (i.e. the sub-bin representing the strongest opacities within a given wavelength bin) at least for solar spectra. This important finding implies that two sub-bins return reasonably accurate results at these wavelengths for the Sun.

Furthermore, the importance of smaller last sub-bins allowed us to optimize the ODF for the cases where speed of calculations is the main concern. A speedup of 3 times compared to the standard method by Kurucz was achieved by using 4 non-uniform sub-bins. In addition to the speedup, our method matches the accuracy of Kurucz longward of 4500\,\AA\ and maintains a comparable level of accuracy shortward. 

For cases where accuracy is the main concern, e.g. shortward of 4500\,\AA\ we find that 10 non-uniform sub-bins are enough to reach a considerably higher level of accuracy compared to the standard sub-bins by Kurucz. That is 2 sub-bins fewer for calculations with the model of the quiet Sun atmosphere.

We also generalized our method for calculations of fluxes through filters typically used for stellar observations. The main idea is to scale the wavelength grid of the detailed opacity by the chosen filter filter transmission curve before the ODF procedure is applied to the changed grid. Note, detailed opacity values are not modified in any way, but the grid over which these opacities are averaged is changed. As an example, this procedure allows the whole Str\"omgren \textit{b} filter to be accurately represented by using only 4 sub-bins. So instead of having many bins for the whole filter, each containing 12 sub-bins, we achieve the same result with just 4, resulting in a speedup of about 2 orders of magnitude. 
The accuracy achieved is better than 0.1\%. 

\begin{acknowledgements}
We thank Remo Collet for encouraging and useful discussions. This work has received funding from the European Research Council under the European Union’s Horizon 2020 research and innovation program (grant agreement No. 715947). It also got a financial support  from the BK21 plus program through the National Research Foundation (NRF) funded by the Ministry of Education of Korea.
\end{acknowledgements}

\bibliographystyle{aa}
\bibliography{lit}
\end{document}